\documentstyle[prb,preprint,aps]{revtex}   
\tightenlines
\begin{document}

\title{
Ginzburg-Landau Theory for a $p$-Wave ${\rm Sr}_2{\rm RuO}_4$ Superconductor: Vortex Core Structure and Extended London Theory 
} 

\author{
R.~Heeb
} 

\address{
Theoretische Physik, ETH H\"onggerberg,
CH-8093 Z\"urich, Switzerland 
} 
\author{ D.F.~Agterberg}
\address{National High Magnetic Field Laboratory, Florida State University, 
Tallahassee, Florida, 32306, USA}
\date{July 1998}
\maketitle
\widetext
\vspace*{-1.0truecm}
\begin{abstract}
Based on a two dimensional odd-parity superconducting order
parameter for Sr$_2$RuO$_4$ with $p$-wave symmetry, we investigate the single vortex and 
vortex lattice structure of the mixed phase near $H_{c1}$. Ginzburg-Landau
calculations for a single vortex show a fourfold structure with an
orientation depending on the microscopic Fermi surface properties.
The corresponding extended London theory is developed to
determine the vortex lattice structure and we find
near  
$H_{c1}$ a centered rectangular vortex lattice. As the field is increased from
$H_{c1}$ this lattice continuously deforms
until a square vortex lattice is achieved. In the centered
rectangular phase
the field distribution, as measurable through $\mu$SR experiments, 
exhibits a characteristic two peak structure 
(similar to that  predicted in high temperature and borocarbide superconductors). 

\end{abstract}


\narrowtext

\section{Introduction}
With the observation of zero resistivity in Sr$_2$RuO$_4$ below $T_c =
0.93 {\rm K}$, Maeno {\it et al.} \cite{maeno94} discovered the first 
layered perovskite compound without CuO$_2$ planes showing
superconductivity. Recent experimental and theoretical
research provides  considerable evidence that the superconducting
state of Sr$_2$RuO$_4$ is of an odd-parity $p$-wave symmetry. 

Classifying the superconducting states of a generalized BCS
theory\cite{sigrist91}, the mean fields 
\begin{equation}
\Delta_{s,s'}(\bbox{k}) = - \sum_{\bbox{k'},\sigma,\sigma'} V_{s' s \sigma
\sigma'}(\bbox{k},\bbox{k'}) \langle c_{\bbox{k'},\sigma} c_{-\bbox{k'},\sigma'}
\rangle
\end{equation}
are split into singlet and triplet pairing states making use of the
antisymmetry condition $\hat{\Delta} (\bbox{k}) = -\hat{\Delta}^{T}
(- \bbox{k})$. While singlet states are described through an even
scalar function $\psi(\bbox{k})$ in the form $\hat{\Delta} (\bbox{k})
= i \hat{\sigma}_y \psi(\bbox{k})$, triplet states take the form
\begin{equation}
\hat{\Delta} (\bbox{k})
\propto i (\bbox{d}(\bbox{k})\cdot \hat{\bbox{\sigma}}) \hat{\sigma}_y,
\end{equation}
where $\bbox{d}(\bbox{k}) = -\bbox{d}(-\bbox{k})$ is an odd vector
function\cite{balian63}. Taking into account the
crystal symmetry, the $\bbox{k}$-dependent functions $\psi(\bbox{k})$
and $\bbox{d}(\bbox{k})$ have to belong to an irreducible
representation of the lattice point group, in our case the tetragonal
point group $D_{4h}$.

Initial proposals of odd parity superconductivity in Sr$_2$RuO$_4$ were founded
on the itinerant ferromagnetism of related ruthenate compounds and on the
similarities between Sr$_2$RuO$_4$ and $^3$He in the Fermi liquid corrections
produced by electronic correlations\cite{rice95}. Recently, muon spin rotation
($\mu$SR) measurements\cite{luke98} have observed that a spontaneous magnetization begins to
develop below $T_c$. This finding is most naturally explained in terms of a time reversal
(${\cal T}$)
symmetry breaking state implying a multiple component representation 
of the $D_{4h}$ point group.
Furthermore, since the electronic structure of Sr$_2$RuO$_4$ is 
quasi-two dimensional, a gap function with a strong $k_z$-dependence is
unlikely.
Of the remaining two odd or even
two-component representations of $D_{4h}$, the odd $\Gamma^-_{5u}$ representation
with basis functions $\{\bbox{d}_1 (\bbox{k}),\bbox{d}_2 (\bbox{k})\} =
\{\hat{\bbox{z}} k_x, \hat{\bbox{z}} k_y \}$ is the only without
nodes at $k_z=0$ and therefore is most likely to be realized in Sr$_2$RuO$_4$. 
 
In Ginzburg-Landau (GL) theory, we deal with order parameters $\eta_j(\bbox{R})$
depending only on the 
center-of-mass coordinate $\bbox{R}$, defined through 
\begin{equation}
\hat{\Delta}(\bbox{R},\bbox{k}) = \sum_{j=1}^2 \eta_j(\bbox{R}) \;i(\bbox{d}_j
(\bbox{k}) \cdot \hat{\bbox{\sigma}}) \hat{\sigma}_y.
\end{equation}
As usual we switch the transformation behavior from the basis functions 
$\bbox{d}_j(\bbox{k})$ to the order parameter components $\eta_j$. The components 
$(\eta_1,\eta_2)$ then share the rotation-inversion symmetry properties of 
($k_x, k_y$),  
and the broken ${\cal T}$-state is $(\eta_1, \eta_2) \propto (1, \pm { i})$.    
Based on this order parameter, one of us\cite{agterberg98-1} examined the  
vortex lattice    
structures of Sr$_2$RuO$_4$ near $H_{c2}$: for an    
applied finite field $H$ along a high symmetry direction in the basal
$ab$-plane, two vortex lattice phases have been found with a  
second superconducting transition between them, while  
a square vortex lattice has minimal energy when the field is along the
$c$-axis.  
Band structure calculations\cite{mackenzie96,oguchi95,singh95}
reveal that the density of states near the Fermi  
surface is mainly due to the four Ru 4$d$ electrons, which hybridize strongly with
the O 2$p$ orbitals and give rise to three bands crossing the Fermi surface, labeled
$\alpha$, $\beta$ and $\gamma$ (see Ref.\onlinecite{mackenzie96}).
Within the model of orbital dependent superconductivity \cite{agterberg97}, the
orientation of the square lattice mentioned above depends on which of the
Fermi surface sheets is 
responsible for the superconductivity. 
Recent experiments using SANS\cite{riseman98} show a   
square vortex lattice aligned with the crystal lattice in a
wide range of the  
phase diagram, indicating that the ${\gamma}$ sheet (due to the $xy$ - Wannier 
functions) of the Fermi
surface shows the superconducting transition\cite{agterberg98-1}. 
Assuming that ferromagnetic spin fluctuations are responsible 
for the superconducting state, this finding is in
agreement with $^{17}$O NMR experiments\cite{imai98} 
showing strong ferromagnetic spin enhancements 
only in the Ru 4$d_{xy}$ orbitals.

In this paper we investigate the single vortex core structure and the geometry
of the vortex lattice near the lower critical field $H_{c1}$. We start from
the Ginzburg-Landau theory for the two-component order parameter
$(\eta_1,\eta_2)$ and solve numerically for the vortex core structure of an
isolated vortex. Next, we derive an extended London model valid for fields near $H_{c1}$ and
large GL parameters $\kappa$ and determine the geometric structure of the
vortex lattice. We find hexagonal lattices near $H_{c1}$,
deforming into square lattices
upon increasing the applied field. Taking the ${\gamma}$ sheet as relevant for
superconductivity, the orientation of the square lattices is in agreement with experiments\cite{riseman98}. The individual vortex cores show square
deformations. This vortex core structure is observable
through scanning tunneling microscopy (STM) measurements since the in-plane 
coherence length of 900 {\AA} is well within the spatial resolution required
by STM.  
Alternatively, Bitter decoration studies near
$H_{c1}$ can be used to detect the predicted field dependent transition between
hexagonal and square lattice, which would complete the picture found by
SANS-measurements for higher applied fields\cite{riseman98}. A further
characteristic of the hex-to-square
transition is the peak splitting in the field distribution calculated in
Sec.~\ref{sec4} and measurable in $\mu$SR experiments. Note that a similar
transition has been predicted 
in borocarbide \cite{kogan96,park98} and high-$T_c$ superconductors 
\cite{affleck97} and has been observed recently
in the borocarbide materials \cite{dewilde97}.

\section{Ginzburg-Landau Results for the Vortex Structure}
\label{glr}
We start from the Ginzburg-Landau free-energy density ($f$) for the
two-dimensional representation $\Gamma_{5u}^-$
of the tetragonal point group, with the
Ginzburg-Landau coefficients determined within a weak-coupling approximation
in the clean limit. 
These two approximations seem reasonable for Sr$_2$RuO$_4$ 
given the ratios\cite{mackenzie98} $T_c/\epsilon_F \sim 10^{-4}$ and $l/\xi \approx 8$ (here
$T_c$ denotes the critical temperature, $\epsilon_F$ the Fermi energy, $l$ the
mean free path and $\xi$ the coherence length). 
In usual dimensionless units, we have\cite{agterberg98-2}

\begin{eqnarray}
\label{glfed}
f & = & -(|\eta_+|^2 + |\eta_-|^2) + \frac{1}{2}\,(|\eta_+|^4+|\eta_-|^4) + 2\, |\eta_+|^2
|\eta_-|^2 + \frac{\nu}{2}\,[(\eta_- \eta_+^*)^2 + (\eta_-^* \eta_+)^2] 
\nonumber\\
& & + B^2 + |\bbox{D} \eta_+|^2
+ |\bbox{D} \eta_-|^2 \nonumber\\
& & 
+\frac{1+\nu}{2}\,[(D_x \eta_+)(D_x \eta_-)^*+(D_x \eta_-)(D_x \eta_+)^*]\nonumber\\
& & 
-\frac{1+\nu}{2}\,[(D_y \eta_+)(D_y \eta_-)^*+(D_y \eta_-)(D_y \eta_+)^*]\nonumber\\
& & 
+\frac{1-\nu}{2\, i}\,[(D_x \eta_+)(D_y \eta_-)^*+(D_y \eta_+)(D_x \eta_-)^*]\nonumber\\
& & 
-\frac{1-\nu}{2\, i}\,[(D_x \eta_+)^*(D_y \eta_-)+(D_y \eta_+)^*(D_x
\eta_-)],
\end{eqnarray}
where $\eta_{\pm} = (\eta_1 \pm i \eta_2)/\sqrt{2}$, $\bbox{B} =
\nabla \wedge \bbox{A}$, $D_\mu = -i \,\nabla_\mu/\kappa -
A_\mu$, $f$ is in units $B_c^2/{4 \pi}$, lengths are in units of the
penetration depth
$\lambda$,
$\bbox{B}$ is in units $\sqrt{2} B_c= \Phi_0/(2\pi\lambda\xi)$,
$\alpha=\alpha_0(T-T_c)$, and
$\kappa=\lambda/\xi$. The anisotropy $\nu = (\langle v_x^4\rangle - 3\langle v_x^2
v_y^2\rangle)/(\langle v_x^4\rangle + \langle v_x^2 v_y^2\rangle)$ is a   
parameter that measures the tetragonal distortion of the Fermi surface 
($\langle \cdot\rangle$ denotes averaging over the Fermi surface; note $|\nu| \le 1$ and $\nu=0$ for a cylindrical geometry). 
The free energy $f$  is invariant under a simultaneous 
rotation of  $45^{\circ}$ around the $c$-axis  
and changing sign of $\nu$. This implies that all results derived below for
$\nu<0$ can be transformed into the corresponding
results for positive $\nu>0$ by a simple rotation, and we limit ourselves to
the case $\nu<0$. 
Further we deal only 
with external fields applied along the $c$-axis allowing us to omit all terms 
containing $z$-derivatives. 

At zero applied field 
the two degenerate solutions $(\eta_+,\eta_-) = (1,0)$ and 
$(\eta_+, \eta_-) = (0,1)$ minimize the free energy $f$. For nonzero applied field, 
the degeneracy is lifted, and     
depending on the direction of the applied field only one component $\eta_\pm$ 
is stable. 
Throughout this paper we will fix the dominant order parameter to be $\eta_-$.
The asymptotic 
vorticity of the dominant $\eta_-$-component then determines the field
direction along
the positive or negative $z$-axis. In the spatially inhomogeneous situations
considered below, the second component $\eta_+$ is 
driven through the mixed
gradient terms [gradient terms such as $(D_x \eta_+)^*(D_y \eta_-)$] 
and decays far away from the defect.
 Note, that to each solution $\{\eta_-,
\eta_+,\bbox{B}\}$ a symmetric counterpart with reversed field direction
exists
\begin{equation}
\{\eta_-, \eta_+, \bbox{B}\}\quad \rightarrow \quad  \{\eta_+^*, \eta_-^*, -\bbox{B}\}.
\end{equation}

Below, we first examine 
the asymptotic solutions of the GL-equations for 
$\nu=0$. Second, we use the $\nu=0$-solution and carry out perturbation theory
in $\nu$ to obtain the asymptotic solutions for $\nu \neq 0$ (we
ignore the coupling to the vector potential in the latter case).

In the cylindrical case ($\nu=0$) the problem exhibits
rotational symmetry. Restricting to fields $\bbox{B} = B(r) \bbox{e_z}$,
the solution has the structure
\begin{eqnarray}
\eta_-(\bbox{r}) & = & e^{i n \theta} \eta_- (r), \nonumber\\
\eta_+(\bbox{r}) & = & e^{i( n+2) \theta} \eta_+ (r), \nonumber\\
\bbox{A}(\bbox{r}) & = & A(r) \bbox{e_{\theta}},
\end{eqnarray}
where $(r,\theta)$ are polar coordinates and $n$ denotes the phase winding of the topological defect  driving
the vortex in the main superconducting
component. The relative phase $e^{i 2 \theta}$  between $\eta_-$ and $\eta_+$
corresponds to an angular momentum difference $\Delta l = 2$ and comes in
through the specific form $(D_x + i D_y)^2$ of the mixed gradient terms for $\nu=0$.
The solutions for $n=\pm 1$ are the most stable ones. The
small $r$ expansion to cubic order takes the form
\begin{eqnarray}
n=1: \qquad
\eta_- (\bbox{r}) & \sim & \left(m_1^0 r -\left(\frac{3}{2} p_3^0 +
\frac{\kappa^2}{8} m_1^0 \left( 1 +
\frac{2}{\kappa} a_1^0\right)\right) r^3\right) e^{i \theta},\nonumber\\
\eta_+ (\bbox{r}) & \sim & p_3^0 r^3\, e^{3 i \theta,} \nonumber\\
A(r) & \sim & a_1^0 r -\frac{(m_1^0)^2}{8 \kappa}\, r^3, \\
n=-1: \qquad  
\eta_- (\bbox{r}) & \sim & \left(m_1^0 r + \frac{\kappa^2}{6} \left(\frac{1}{2} p_1^0 - m_1^0 +
\frac{3}{\kappa} a_1^0 \left(m_1^0+p_1^0\right)\right) r^3\right) e^{-i \theta},\nonumber\\
\eta_+ (\bbox{r}) & \sim & \left(p_1^0 r + \frac{\kappa^2}{6} \left(\frac{1}{2} m_1^0 - p_1^0 -
\frac{3}{\kappa} a_1^0 \left(m_1^0+p_1^0\right)\right) r^3\right) e^{i \theta}, \nonumber\\
A(r) & \sim & a_1^0 r -\frac{(p_1^0)^2-(m_1^0)^2}{8 \kappa}\, r^3,
\end{eqnarray}
where $m_1^0$, $p^0_{1,3}$, and $a_1^0$ are the first expansion coefficients 
of $\eta_-$, $\eta_+$, and $A$, to be determined by matching 
this small $r$ expansion to the asymptotics at $r \rightarrow \infty$. The
latter reads
\begin{eqnarray}
\eta_-(\bbox{r}) & = & (1 - m(r))\, e^{i n \theta},  \nonumber\\
\eta_+(\bbox{r}) & = & p(r)\, e^{i (n+2) \theta}, \nonumber\\
A(r) & = & \pm \frac{1}{\kappa r} + a(r),
\end{eqnarray}
where the $\pm$ stands for $n=\pm 1$. The asymptotic form of the GL-equations
requires the three functions $m(r)$, $p(r)$, and $a(r)$ to be of the form
\begin{eqnarray}
\eta_- (\bbox{r}) & \sim & \left(1- \frac{(n+2)^2}{2 \kappa^2 r^2}\right)
e^{i n \theta},\nonumber\\
\eta_+ (\bbox{r}) & \sim & \frac{n^2+2n}{2 \kappa^2 r^2}\,
e^{i(n+2) \theta},
\end{eqnarray}
if we ignore the vector potential $A(r) = 0$. For the general case
($A(r) \neq 0$) the main asymptotics is an exponential decay
\begin{equation}
a(r), m(r), p(r) \sim  e^{- r/\Lambda},
\end{equation}
which is a consequence of transverse screening.
From Fig.~\ref{expofig} we
see, that for reasonably large $\kappa$ this
exponent $\Lambda$ approaches 1, which equals the usual
penetration depth in dimensionless units.

We proceed now with the anisotropic case
$\nu \not= 0$ (here we neglect the vector potential). 
An expansion to first order in $\nu$ produces the following result for
small $r$
\begin{eqnarray}
n=1: \qquad
\eta_- (\bbox{r}) & \sim & \left(m_1^0 r -\frac{\kappa^2 m_1^0 + 12 p_3^0}{8}\, r^3\right)
e^{i \theta} + \nu \left(\frac{\kappa^2 m_1^0 + 12 p_3^0}{16} - \frac{3 g^0_{-,3}}{2}\right)\,
r^3 e^{-i \theta}, \nonumber\\
\eta_+ (\bbox{r}) & \sim & p_3^0 r^3 \,e^{3 i \theta} + \nu g^0_{-,3} r^3\, e^{-i \theta},\\ 
n=-1: \qquad
\eta_- (\bbox{r}) & \sim & \left(m_1^0 r +\frac{\kappa^2}{12} (p_1^0-2 m_1^0) r^3\right)
e^{-i \theta} + \nu g^0_{+,3} r^3\, e^{3 i \theta}, \nonumber\\
\eta_+ (\bbox{r}) & \sim & \left(p_1^0 r +\frac{\kappa^2}{12} (m_1^0-2 p_1^0) r^3\right)
e^{i \theta} + \nu f^0_{-,3} r^3\, e^{-3 i \theta}, 
\end{eqnarray}
and for large $r$ 
\begin{eqnarray}
\eta_- (\bbox{r}) & \sim & \left(1- \frac{(n+2)^2}{2 \kappa^2 r^2}\right)
e^{i n \theta} - \nu \frac{n}{2 \kappa^2 r^2} \left(e^{i(n+4)\theta} +3 e^{i (n-4)\theta} \right),\nonumber\\
\eta_+ (\bbox{r}) & \sim & \frac{n^2 + 2n}{2 \kappa^2 r^2}\,
e^{i(n+2) \theta} + \nu \frac{n}{\kappa^2 r^2} \left(e^{i(n+6)\theta} + e^{{
i} (n-2)\theta} \right).
\end{eqnarray}
The additional coefficients $g^0_{\pm,3}$ and $f^0_{\pm,3}$ are the first
expansion coefficients of the higher angular momentum states ($f^0_{\pm,3}$
from $\eta_+ \sim e^{i (n+2 \pm 4) \theta}$ and $f^0_{\pm,3}$
from $\eta_- \sim e^{i (n \pm 4) \theta}$), which arise
for $\nu \neq 0$.
Comparing the condensation energy for the two cases $n=\pm1$ we find that the $(n=-1)$-case is the
more stable: the admixed $\eta_+$ order parameter rises linearly
in the center rather than cubic as is the case for $n=1$, 
and the asymptotic phase turns of
$\eta_+$ are the same in the center and for large $r$ (which is also an argument
against off-centered zeroes as seen for example in
$d_{x^2-y^2}$-superconductors). Both points lead to a higher gain in
condensation energy, which should  stabilize the $(n=-1)$-solution. 
Our numerical
Ginzburg-Landau results (described in the next paragraph) for the
critical fields $H_{c1}=\epsilon_l \kappa/4\pi$ confirm this finding: for $\kappa=2.5$ and
$\nu=-0.3$, we obtain
\begin{eqnarray*}
H_{c1}(n=-1) & = & 0.43 B_c,\\
H_{c1}(n=1) & = & 0.47 B_c ,
\end{eqnarray*}
demonstrating the stability of the $(n=-1)$-solution compared to $(n=1)$.
The corresponding $s$-wave result is $H_{c1}=0.48 B_c$.

To obtain complete numerical solutions, we minimize the free energy density
on a two-dimensional grid. The minimization is carried out by  
relaxing an initial configuration $\eta_+=0$, $\eta_-=\tanh(r) \,e^{\pm i \theta}$ and $\bbox{A}=0$ 
iteratively. Indeed we find that the energetically
favorable $(n=-1)$-solution shows no off-centered zeroes in $\eta_+$.
A contour plot of the absolute value of the dominant
$\eta_-$ (a) and $\eta_+$-component (b) is displayed in
Fig.~\ref{eta-GLfig}. The admixed $\eta_+$ solution
shows a fourfold structure with a maximal amplitude along the
axes. In addition the
vortex core center of the dominant
order parameter is distorted with marked elongations along the diagonals
of the crystal lattice for $\nu<0$ (and
analogously along the axes for $\nu>0$). This behavior is understood
recalling the properties of the Fermi surface. For $\nu<0$,
the density of states (DOS) at the Fermi surface is larger along the axes
than along the diagonals. A larger DOS however implies also a larger effective
mass, which reduces the coherence length and thus the
size of the vortex.
Finally, the distribution of the absolute value 
of the $B_z$-field (see Fig.~\ref{A-GLfig}) also shows clear square 
deformations aligned with the underlying crystal lattice. 

On the other hand, in the unstable $(n=1)$-case the admixed component
$\eta_+$ exhibits a negative phase turn around the center, and consequently four
off-centered zeroes with positive phase turn are needed to match up this phase
$e^{-i \theta}$ at small $r$ to the asymptotic phase $e^{3 i \theta}$ at large
$r$. This pattern is qualitatively well known from analog GL-calculations for $s$-admixture in
$d_{x^2-y^2}$  
high-$T_c$
superconductors\cite{heeb96,soininen94,berlinsky95,franz96,ichioka96}. The
dominant component $\eta_-$ and the $B_z$-field show similar features as in
the stable case discussed above.

\section{Extended London Theory}
\label{exlond}
We wish to determine the form of the vortex lattice  
close to $H_{c1}$.
Instead of solving the Ginzburg-Landau free energy directly, here we follow
an idea recently developed in the context of 
$d_{x^2-y^2}$-wave
superconductors with a weak admixing of an $s$-component\cite{affleck97}. We
first integrate 
out the admixed component, leading to a one-component free energy
density. We assume $\kappa \gg 1$ and $\nu \ll 1$ and use London approximation to obtain an
effective free energy density which depends only on the field $B$ and which can be
minimized with respect to the
geometry of the vortex lattice. 
Though our results will be quantitatively inaccurate for the low-$\kappa$
Sr$_2$RuO$_4$ material, we still expect to obtain a qualitatively correct picture.

The Ginzburg-Landau equation for $\eta_+$ to first order in $|\eta_+| /|\eta_-|$ reads
\begin{equation}
0 = -\eta_+ + 2\, |\eta_- |^2 \eta_+ + {\bbox{D}}^2 \,\eta_+ + \frac{1+\nu}{2}\, (D_x^2 - D_y^2) \,\eta_- -
\frac{1-\nu}{2\, i}\, (D_x D_y + D_y D_x)\, \eta_- .
\end{equation}
Within London theory the dominant component $\eta_-$ has modulus unity (see
Eq.~(\ref{asf})) and the first three terms lead to the expression
$(1+{\bbox{D}}^2) \,  \eta_+$.
Solving formally for $\eta_+$ and substituting this expression into the free energy density,
we obtain the effective  one-component Ginzburg-Landau free energy 
\begin{equation} 
f =  -|\eta_-|^2 + \frac{1}{2}\,|\eta_-|^4 
+ |\bbox{D} \eta_-|^2 - \eta_-^* \, {\cal P}(\bbox{D},\nu) \left(1+{\bbox{D}}^2
\right)^{-1} {\cal P}(\bbox{D},\nu)\, \eta_- + B^2,
\end{equation}
with 
\begin{equation}
{\cal P}(\bbox{D},\nu) =  \frac{1+\nu}{2}\, (D_x^2-D_y^2) -
\frac{1-\nu}{2 {\,i}} \, (D_x
D_y+ D_y D_x). 
\end{equation}
Here we have consistently neglected all terms of order $\eta_+^4$,
$\nu \eta_+
^2$, and higher. 

Following the usual scheme for London approximation we replace the order
parameter $\eta_-$ and its derivative by the superfluid velocity $\bbox{v}$
\begin{eqnarray} 
\label{asf}
\eta_- & = & e^{ i \phi(\bbox{x})},\nonumber\\
\bbox{D} \eta_- & = & ( \frac{1}{\kappa}\, \bbox{\nabla} \phi(\bbox{x}) - \bbox{A})
\,e^{ i
\phi(\bbox{x})} \equiv
 \frac{1}{\kappa} \,\bbox{v} \,e^{ i \phi(\bbox{x})}.
\end{eqnarray}
Keeping only terms up to second order in $\bbox{v}$
and up to first order in the anisotropy parameter $\nu$,
we obtain the London free energy in Fourier representation
\begin{eqnarray}
\label{londfed}
f(\bbox{k}) & = & -\frac{1}{2}+\frac{1}{\kappa^2}\, \bbox{v}^2 + B^2
\nonumber\\
& & - \frac{1}{4 \kappa^4}\; \frac{1}{1+k^2/\kappa^2}\,
\Big(  k^2 \bbox{v}^2 + 2 \nu
\left[(k_x^2 - k_y^2)\, (v_x^2 - v_y^2) - (2\, k_x k_y)\, (2\, v_x v_y)
\right] \Big).
\end{eqnarray}
Here, all quadratic expressions in $\bbox{v}$ and $B$ have to be understood as
$v_i v_j = v_i(\bbox{k}) v_j(-\bbox{k})$ (the results for vorticity $n=\pm 1$ are the same on this level of
calculation; they would differ if one included higher orders in $\bbox{v}$).

Next, we eliminate $\bbox{v}$. We obtain an equation connecting $B$ and
$\bbox{v}$ by variation of the London free energy Eq.~(\ref{londfed}) with respect to the
vector potential $\bbox{A}$. The corresponding Euler-Lagrange equation 
reads
\begin{equation}
i\, \kappa^3 \:\bbox{k} \wedge \bbox{B}  =  \kappa^2\, \bbox{v} 
- \frac{1}{4}\; \frac{1}{1+k^2/\kappa^2}\, \Big( k^2 \bbox{v} + 2 \nu \left[ (k_x^2-k_y^2)\, (v_x \bbox{\hat x} - v_y \bbox{\hat y})
- (2\, k_x k_y)\, (v_x \bbox{\hat y} + v_y \bbox{\hat x} ) \right] \Big).
\end{equation}
We solve this equation for $\bbox{v}$ up to first order in $\nu$ and
substitute the solution back into the London
free energy density (Eq.~(\ref{londfed})), to obtain
\begin{eqnarray}
f(\bbox{q}) & = & -\frac{1}{2} + \left[ 1 + q^2 \kappa^2 \frac{1+q^2}{(1+\frac{3}{4}\, q^2)^2}
\left(1 + \frac{3}{4}\, q^2 - \frac{\nu}{2} \Big(\frac{(q_x^2 - q_y^2)^2}{q^2} - \frac{(2\, q_x
q_y)^2}{q^2} \Big) \right) \right] B^2 \nonumber \\
& \equiv & -\frac{1}{2} + J(\bbox{q},\nu,\kappa)\, B^2,
\end{eqnarray}
where we introduced the short notation 
\begin{equation}
\bbox{q} = \frac{1}{\kappa}\, \bbox{k}.
\end{equation}
The variation with respect to $B(\bbox{k})$ leads to the extended London 
equation $J(\bbox{q},\nu,\kappa) B({\bbox{k}}) =0$ (in the limit $\kappa
\rightarrow \infty$ we recover the
well-known result $(1+k^2) B({\bbox{k}})=0$ (no sources)).

\section{Vortex Lattice close to $H_{\lowercase{c}1}$}
\label{sec4}
We generate the vortex lattice  by
introducing the corresponding source terms at the vortex core positions into
the London
equations. A convenient
form for these source terms is\cite{brandt77}
\begin{equation}
\sigma(\bbox{r}) = \kappa \sum_j {e^{-{\kappa^2 (\bbox{r} -\bbox{r}_j)^2}/{2}}},
\end{equation}
where the sum leads over all positions $\bbox{r}_j$ of the vortex lattice. The
magnetic field and the Gibbs free energy 
for the vortex lattice read
\begin{eqnarray}
\label{gibbs}
B(\bbox{r}) & = & {\overline B} \, \sum_{\bbox{q}_j}\, \frac{e^{i \kappa \bbox{q_j}\bbox{r}}\, e^{- {q_j^2}/{2}}}{J(\bbox{q}_j,\nu,\kappa)} \nonumber\\
{\cal G}(H) & = & {\overline B}^2 \,  \sum_{\bbox{q}_j} \frac{e^{-
{q_j^2}}}{J(\bbox{q}_j,\nu,\kappa)} - 2 H {\overline B},
\end{eqnarray}
where the specific form of the vortex lattice is given by the reciprocal
lattice vectors $\bbox{q}_j$. We  parameterize the unit
cell in real space through the angles  $\theta_1$ and $\theta_2$, the ratio of
the lengths of the unit vectors $\zeta$ and the average field ${\overline B}$ (see
Fig.~\ref{unitcellfig}), where the first three parameters describe the geometry
and ${\overline B}$ the size of the cell. The lattice geometry depends on the
applied field $H$ and is determined by minimizing the Gibbs free energy ${\cal
G}(H)$ given in Eq.~(\ref{gibbs}) with respect to the unit cell parameters.
Writing the 
operator $J(\bbox{q},\nu,\kappa)$ in the form
\begin{equation}
J(\bbox{q},\nu,\kappa) = 1 + \kappa^2 \frac{1+q^2}{(1+\frac{3}{4}\, q^2)^2}
\left(q^2 + \frac{3}{4}\, q^4 - \frac{\nu}{2} \, q^4 \cos {4 \phi_{\bbox{q}}} \right),
\end{equation}
it is obvious that a
change of sign in $\nu$ still  corresponds to a rotation in real and
reciprocal space
by $45^\circ$. We thus restrict the discussion again to negative values of $\nu$.
The resulting vortex lattices exhibit a centered rectangular unit
cell with the main axes aligned along the diagonals of the underlying
crystal. Without loss of generality we thus assume $\theta_1 =
\pi/4 -\theta_2/2$ and $\zeta = 1$
in the remainder of the paper. 
In Fig.~\ref{bmin-Hfig}  we plot the angle $\theta_2$ minimizing the
Gibbs free energy for different values of $\nu$ versus the field $H$
(Fig.~\ref{bmin-Hfig}a: $\kappa = 25.0$, Fig.~\ref{bmin-Hfig}b: $\kappa = 5.0$). For large enough values of the anisotropy parameter $\nu$, the vortex
lattice evolves from hexagonal at low applied fields to square
at high fields. The lattices found in our calculations ($\nu
<0$) are
always aligned with the underlying lattice (see
Fig.~\ref{contfig}). For decreasing $\kappa$, the
crossover to a square  lattice  approaches to the
lower critical
field $H_{c1}$ and occurs at lower anisotropies $\nu$, an effect which is believed to be even stronger if one would
include all nonlinearities into the theory.
These results are in good agreement with recent experimental data of Riseman
et al.~\cite{riseman98}, who find an aligned square lattice in Sr$_2$RuO$_4$ down to very small
fields in the phase diagram.

Apart form the vortex lattice symmetry, the shape of the individual vortices
as well as the corresponding field distributions is of interest. The
former, because STM-measurements are able to determine
the form of the individual vortices directly, and the latter since the field
distribution is the typical result of $\mu$SR-measurements. While the shape of
the 
$B$-field around the vortex core is perfectly circular for $\nu = 0$ over the
whole range of the applied field $H$, we find square deformations
of the $B$-field
for negative values of $\nu$ (see Fig.\ref{contfig}). This is in agreement with
the
single vortex results obtained in the previous section, which also exhibits the
corresponding deformation of the fields around the vortex core. 

Experiments using $\mu$SR can observe the field distribution
\begin{equation}
P(B) = \frac{1}{V_{\rm uc}} \int_{V_{\rm uc}} d^2 \bbox{r}\: \delta\,(B-B(\bbox{r})).
\end{equation}
In Fig.~\ref{ph-Hfig},
characteristic results are shown for two cases $\kappa = 25.0, H = 3.1$ (a)
and $\kappa = 5.0, H = 0.5$ (b). While at $\nu = 0$ the field
distribution shows only one peak with a small shoulder towards the minimal
field value, the situation for $\nu \not= 0$ is quite different. With
increasing $|\nu|$ the $\nu=0$-peak splits into
two due to the appearance of two nonequivalent classes of saddle points in the
$B$-field. Upon further increase of $|\nu|$, the low field peak gradually vanishes and finally develops into a broad
shoulder. 

\section{Conclusions}
\label{conc}
We have investigated the single vortex and vortex lattice structure
of Sr$_2$RuO$_4$ near $H_{c1}$, assuming a weak-coupling, clean limit
description of the 2D odd-parity $\Gamma_{5u}^-$ symmetry of the superconducting state in Sr$_2$RuO$_4$.
The stable single vortex solution (denoted $(1,-1)$)
takes the asymptotic form
$(\eta_+,\eta_-)=(p(r)e^{i \theta},[1-m(r)]e^{-i \theta})$. A second low energy
vortex solution has the asymptotic form  
$(\eta_+,\eta_-)=(p(r)e^{3i \theta},[1-m(r)]e^{i \theta})$ (denoted
$(3,1)$). Solutions of the latter type appear to be the minimal ones for $d$-wave superconductors with 
an admixed $s$-wave component. Here, the symmetry allows for the competing
$(1,-1)$-solution which has a lower energy as it contains fewer nodes 
and decays slower near the vortex core. 
The stable $(1,-1)$-solution shows clear four-fold deformations due
to the anisotropy of the density of states at the Fermi surface. Current
experimental results on the vortex lattice structure point towards an elongation
of the vortex along the crystal lattice diagonals, in agreement with our
theoretical analysis. A direct proof of such a structure
should be observable through STM measurements.

The structure of the vortex lattice has been determined using an extended London
approach where the admixed $\eta_+$ component has been integrated out. 
The resulting
London theory predicts 
that with increasing field the vortex lattice will continuously deform
from a hexagonal to a square lattice. The orientation
of the square
vortex lattice depends on the anisotropy parameter $\nu$:
For $\nu<0$ the square vortex lattice  is aligned with the 
crystal lattice 
whereas for $\nu>0$ the orientation of the vortex lattice
is rotated by $45^{\circ}$. Comparing to recent experimental results of
Riseman {\it et al.}\cite{riseman98}, who found the vortex and crystal
lattices aligned in almost the
entire field range, we conclude that $\nu<0$. The continuous hex-to-square
transition should be observable through Bitter decoration experiments. 
An alternative signature of this transition is given through the
splitting in the peak of the field distribution $P(B)$. These field distributions
should be observable in $\mu$SR experiments.

\subsection*{Acknowledgments}
We thank G. Blatter, E.~M. Forgan, G. Luke, A. Mackenzie, 
Y. Maeno, T.~M. Rice, and  especially 
M. Sigrist for helpful discussions. We
gratefully acknowledge support by the Swiss Nationalfonds 
and D.~F.~A. further acknowledges support from the Natural Science and Engineering
Research Council of Canada.

\begin{figure}
\caption{Exponent $\Lambda$ governing the decay of $\nu=0$ solutions.}
\label{expofig}
\end{figure}

\begin{figure}
\caption{Contour plot of GL-calculations for the absolute values of the dominant
$\eta_-$- (a) and the admixed $\eta_+$-component (b) for the parameters
$\kappa = 2.5$ and $\nu=-0.3$. The contours are $0.99$,
$0.975$, $\ldots$ for (a) and $0.03$,
$0.045$, $\ldots$, 0.225 for (b)}
\label{eta-GLfig}
\end{figure}

\begin{figure}
\caption{Contour plot of the absolute value of the $|B|$-field for the
same parameters as in Fig.~\ref{eta-GLfig}. Notice the strong square deformations of the 
vortex core along the diagonals of the crystal lattice. The contours are $0.03$,
$0.06$, $\ldots$, 0.42. }
\label{A-GLfig}
\end{figure}

\begin{figure}
\caption{Schematic illustration of the real unit cell.}
\label{unitcellfig}
\end{figure}

\begin{figure}
\caption{Vortex lattice structure as function of the applied field
$H$. The GL-parameter is $\kappa=2.5$ (a) and
$\kappa = 25.0$ (b), and $\theta_2$ denotes the angle between the unit cell
vectors minimizing the free 
energy density given in Eq.~(\ref{gibbs}). }
\label{bmin-Hfig}
\end{figure}

\begin{figure}
\caption{Contour plot of the field distribution for
$\nu = -0.4, \kappa = 2.5, H = 0.25$. The horizontal and vertical axes of the
plot correspond to the axes of the underlying crystal
lattice. The field distribution shows an aligned square lattice with weak
square shaped deformations of the vortex cores.}
\label{contfig}
\end{figure}

\begin{figure}
\caption{Field distribution $P(B)$ for different values of the anisotropy $\nu$ at
$\kappa = 2.5, H = 0.3$ (a) and $\kappa = 25.0, H = 3.1$ (b).}
\label{ph-Hfig}
\end{figure}


\begin{thebibliography}{10}

\bibitem{maeno94}
Y.~Maeno, H.~Hashimoto, K.~Yoshida. S.~Nishizaki, T.~Fujita, J.~G.~Bednorz,
and F.~Lichtenberg,
\newblock Nature {\bf 372}, 532 (1994).

\bibitem{sigrist91}
M.~Sigrist and K.~Ueda,
\newblock Rev.\ Mod.\ Phys. {\bf 63}, 239 (1991).

\bibitem{balian63}
T.~Balian and N.~R.~Werthamer,
\newblock Phys.\ Rev.\ {\bf 131}, 1553 (1963).

\bibitem{rice95}
T.~M.~Rice and M.~Sigrist,
\newblock J.\ Phys.\ Condens.\ Matter {\bf 7}, L643 (1995).

\bibitem{luke98}
G.~M.~Luke, Y.~Fudamoto, K.~M.~Kojima, M.~I.~Larkin, J.~Merrin, B.~Nachumi,
Y.~J.~Uemura, Y.~Maeno, Z.~Q.~Mao, Y.~Mori, H.~Nakamura, and M.~Sigrist,
\newblock Nature (London) {\bf 394}, 558 (1998).

\bibitem{agterberg98-1}
D.~F.~Agterberg,
\newblock Phys.\ Rev.\ Lett. {\bf 80}, 5184 (1998).

\bibitem{mackenzie96}
A.~P.~Mackenzie, S.~R.~Julian, A.~J.~Diver, G.~G.~Lonzarich, Y.~Maeno, S.~
Nishizaki, and T.~Fujita,
\newblock Phys.\ Rev.\ Lett. {\bf 76}, 3786 (1996).

\bibitem{oguchi95}
T.~Oguchi,
\newblock Phys.\ Rev.\ B {\bf 51}, 1385 (1995).

\bibitem{singh95}
D.~J.~Singh,
\newblock Phys.\ Rev.\ B {\bf 52}, 1358 (1995).

\bibitem{agterberg97} 
D.~F.~Agterberg, T.~M.~Rice, and M.~Sigrist,
\newblock Phys.\ Rev.\ Lett.\ {\bf 78}, 3374 (1997).

\bibitem{riseman98}
T.~M.~Riseman, P.~G.~Kealy, E.~M.~Forgan, A.~P.~Mackenzie, L.~M.~Galvin,
A.~W.~Tyler, S.~L.~Lee, C.~Ager, D.~McK.~Paul, C.~M.~Aegerter, R.~Cubitt, 
Z.~Q.~Mao, S.~Akima, and Y.~Maeno, 
\newblock Nature  (London), in press.

\bibitem{imai98}
T.~Imai, A.~W.~Hunt, K.~R.~Thurber, and F.~C.~Chou,
\newblock Phys. Rev. Lett.  {\bf 81}, 3006  (1998).

\bibitem{kogan96} 
V.~G.~Kogan, A.~Gurevich, J.~H.~Cho, D.~C.~Johnston, Ming
Xu, J.~R.~Thompson, and A.~Martynovich,
\newblock Phys.\ Rev.\ B {\bf 54}, 12386 (1996).

\bibitem{park98} K.~Park and D.~A.~Huse,
\newblock Phys.\ Rev.\ B {\bf 58}, 9427 (1998).

\bibitem{affleck97}
I.~Affleck, M.~Franz, and M.~H.~S.~Amin,
\newblock Phys.\ Rev.\ B {\bf 55}, R704 (1997).

\bibitem{dewilde97} Y.~De Wilde, M.~Iavarone, U.~Welp, V.~Metlushko,
A.~E.~Koshelev, I.~Aranson, G.~W.~Crabtree, and P.~C.~Canfield,
\newblock Phys.\ Rev.\ Lett.\ {\bf 78}, 4273 (1997).

\bibitem{mackenzie98}
A.~P.~Mackenzie, R.~K.~W.~Haselwimmer, A.~W.~Tyler, G.~G.~Lonzarich, Y.~Mori, S.~
Nishizaki, and Y.~Maeno,
\newblock Phys.\ Rev.\ Lett.\ {\bf 80}, 161 (1998).

\bibitem{agterberg98-2}
D.~F.~Agterberg,
\newblock Phys.\ Rev.\ B, to appear Dec. 1, 1998.

\bibitem{heeb96}
R.~Heeb, A.~van\ Otterlo, M.~Sigrist, and G.~Blatter,
\newblock Phys.\ Rev.\ B {\bf 54}, 9385 (1996).

\bibitem{soininen94}
P.~I.~Soininen, C.~Kallin, and A.~J.~Berlinsky,
\newblock Phys.\ Rev.\ B {\bf 50}, 13883 (1994).

\bibitem{berlinsky95}
A.~J.~Berlinsky, A.~L.~Fetter, M.~Franz, C.~Kallin, and P.~I.~Soininen,
\newblock Phys.\ Rev.\ Lett. {\bf 75}, 2200 (1995).

\bibitem{franz96}
M.~Franz, C.~Kallin, P.~I.~Soininen, A.~J.~Berlinsky, and A.~L.~Fetter,
\newblock Phys.\ Rev.\ B {\bf 53}, 5795 (1996).

\bibitem{ichioka96}
M.~Ichioka, N.~Hayashi, N.~Enomoto, and K.~Machida,
\newblock Phys.\ Rev.\ B {\bf 53}, 2233 (1996).

\bibitem{brandt77}
E.~H.~Brandt,
\newblock J.\ Low Temp.\ Phys. {\bf 26}, 709, 735 (1977).

\end{thebibliography}
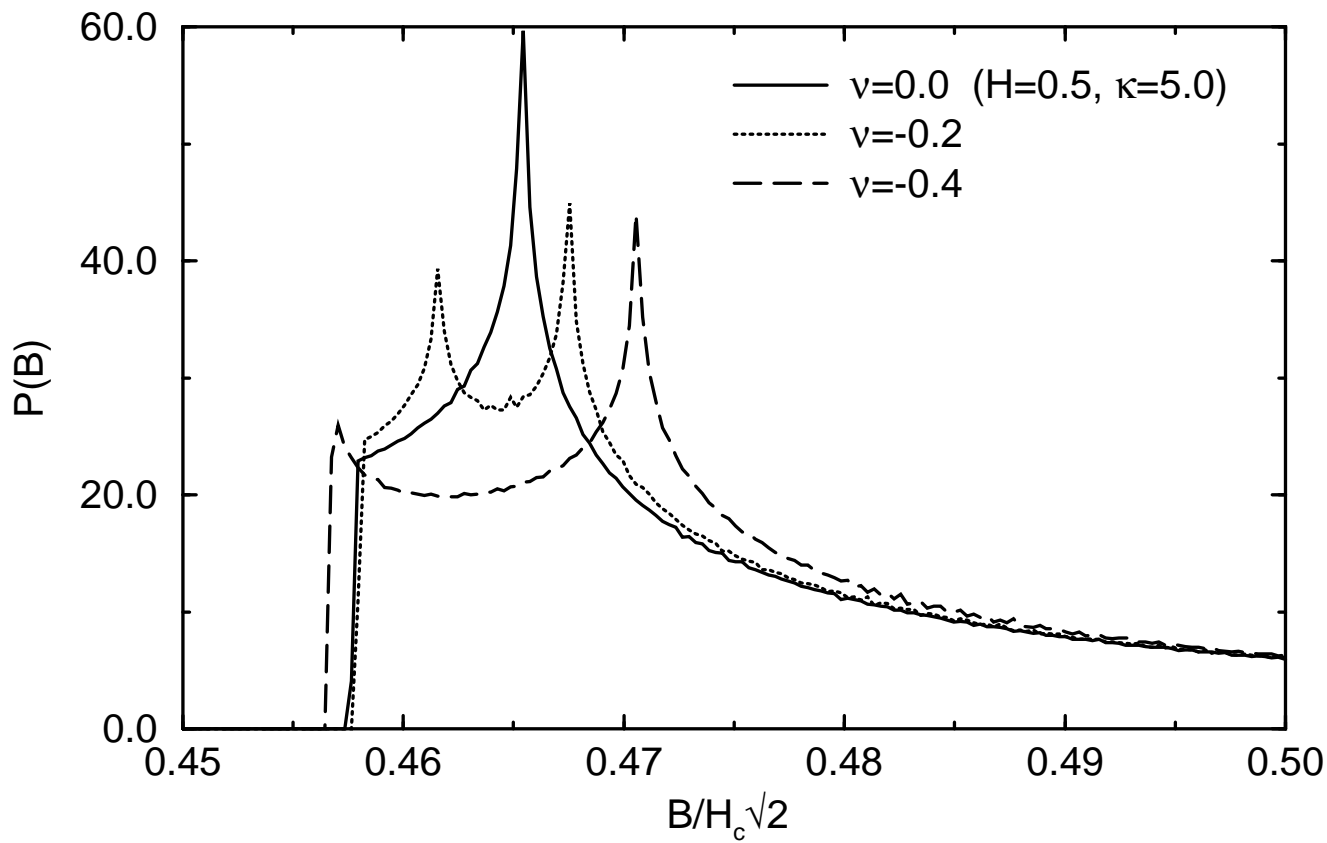
\end{document}